\documentclass[10pt,letterpaper]{article}
\usepackage[top=0.85in,left=2.75in,footskip=0.75in]{geometry}

\usepackage{amsmath,amssymb}

\usepackage{changepage}

\usepackage[utf8x]{inputenc}

\usepackage{textcomp,marvosym}

\usepackage{cite}

\usepackage{nameref,hyperref}

\usepackage[right]{lineno}

\usepackage{microtype}
\DisableLigatures[f]{encoding = *, family = * }

\usepackage[table]{xcolor}

\usepackage{array}

\newcolumntype{+}{!{\vrule width 2pt}}

\newlength\savedwidth

\raggedright
\usepackage[aboveskip=1pt,labelfont=bf,labelsep=period,justification=raggedright,singlelinecheck=off]{caption}

\makeatletter
\renewcommand{\@biblabel}[1]{\quad#1.}
\makeatother

\usepackage{lastpage,fancyhdr,graphicx}
\usepackage{epstopdf}
\fancyheadoffset[L]{2.25in}
\fancyfootoffset[L]{2.25in}
\begin{document}
\vspace*{0.2in}

\begin{flushleft}
{\Large
\textbf\newline{Thermodynamics of harmony: extending the analogy across musical systems.} }
\\
L. Nasser$^{*}$\textsuperscript{1\Yinyang\*},
A. Tillotson\textsuperscript{2\Yinyang},
X. Hernandez\textsuperscript{3\Yinyang}

\bigskip
\textbf{1} Dept of Science and Mathematics, Columbia College Chicago, Chicago, IL, USA
\\
\textbf{2} Dept of Physics, NYU, Abu Dhabi, UAE
\\
\textbf{3} Instituto de Astronom\'{\i}a, Universidad Nacional Autónoma de México, Apartado Postal 70-264 C.P. 04510 México, D.F. México
\\
\bigskip

\Yinyang These authors contributed equally to this work.

* lnasser@colum.edu

\end{flushleft}
% Please keep the abstract below 300 words
\section*{Abstract} It is common for most people to think of science and art as disparate, or at most only vaguely related fields. In physics, one of the biggest successes of thermodynamics is its explanation of order arising from disordered phases of matter through the minimization of free energy; In 2019, Berezovsky showed \cite{bib1} that the mechanism describing emergent order from disorder in matter can be used to explain how ordered sets of pitches can arise out of disordered sound, thus bridging the gap between science and the arts in a powerful way. In this paper we analyze his method in detail, generalizing it beyond the 12 tone system of intonation of Western music by explicitly considering Gamelan instruments and clarifying some details in the hope of strengthening it and making it better known and recognized.

% Please keep the Author Summary between 150 and 200 words
% Use first person. PLOS ONE authors please skip this step.
% Author Summary not valid for PLOS ONE submissions.  
%\section*{Author summary}
%Lorem ipsum dolor sit amet, consectetur adipiscing elit. Curabitur eget porta erat. Morbi consectetur est vel gravida pretium. Suspendisse ut dui eu ante cursus gravida non sed sem. Nullam sapien tellus, commodo id velit id, eleifend volutpat quam. Phasellus mauris velit, dapibus finibus elementum vel, pulvinar non tellus. Nunc pellentesque pretium diam, quis maximus dolor faucibus id. Nunc convallis sodales ante, ut ullamcorper est egestas vitae. Nam sit amet enim ultrices, ultrices elit pulvinar, volutpat risus.

%\linenumbers

% Use "Eq" instead of "Equation" for equation citations.
\section*{Introduction}
In Shoenberg's classic treatise ``Theory of Harmony'' \cite{bib3}, he states ``Art in its most primitive state is a simple imitation of nature. But it quickly becomes imitation of nature in the wider sense of this idea, that is, not merely imitation of outer but also of inner nature''. Provocative as that statement is, it's unlikely Schoenberg ever imagined that one day it would be possible to trace the origins of harmony to thermodynamics by direct analogy to the mechanisms that give rise to ordered phases in nature. To our knowledge, the first investigation in this direction was by Berezovsky \cite{bib1} \cite{bib8}. In this paper we extend the ideas explored in \cite{bib1}, clarifying and exploring in detail what are the essential elements needed for the method to work and generalizing some of the results needed to understand harmony as the result of an ordered phase that arises from interacting sounds. Before we outline how this was done, it is useful to review some basic ideas of music theory.

Harmony refers to the sound of two or more notes played together. Its rules are based on certain relationships among notes that the human ear will either intuitively accept or reject. These rules are also expressible mathematically and have been the subject of scientific investigation, dating back as far as the 6th Century BCE. with Pythagoras. He famously showed that by comparing the sound made by plucking strings of different lengths, the distances between notes (or musical \textit{intervals}) that the ear found agreeable had lengths in specific ratios, and is said to have exclaimed: ``There is geometry in the humming of the strings.'' Examples of these intervals are the \textit{octave} corresponding to lengths in the ratio $(1:2)$, the \textit{perfect fifth} for lengths in the ratio $(2:3)$, the \textit{fourth} for lengths in the ratio $(3:4)$, and so forth. In essence, Western harmony is built upon chords that are purposefully constructed upwards from their bass or \textit{fundamental} note, using other notes whose intervals are perceived as agreeable; a succession of chords is then analyzed by the distance, or intervals, between their roots. 

The use of the word ``agreeable'' is worthy of note: By necessity it requires human perception which by no means is a mathematically exact standard. Indeed, music can be seen as a language that employs the subtle interplay between the perception of consonance and dissonance to convey an emotional response in the listener. Consonance refers to the combination of notes accepted as ``agreeable'' or restful. Dissonance refers to combinations of notes that are perceived as tense, and it is a key element in music; dissonance creates movement and gives the music flow between states of tension and relaxation as the dissonance is followed by consonance; without dissonance, music would be static and boring. Given its crucial role, it is essential to bear in mind that dissonance is a culturally-shaped construct, and thus, it helps us understand why different cultures employ different musical styles and structures to express themselves.

The octave is of very special interest in the construction of a musical system. In physical terms, an octave is an interval defined by a note of fundamental frequency $f$ and another with fundamental frequency $2f$. These two notes are essentially perceived as ``the same note'', differing only by their rate of vibrations which manifests in the perception of one being ``higher'' than the other \cite{bib11}. A scale is a succession of notes that form a progression from a note to its octave. In Western music, it was established that the octave should be divided into 12 steps. Initially the size of the interval between successive notes was not constant, and this led to problems with intonation \cite{bib7},\cite{bib14} that forbade polyphony. To understand why, consider the way in which Pythagoras constructed the scale \cite{bib7}: he took a base note of frequency $f$, and raised it by a fifth, which means multiplying it by $\frac{3}{2}$. You then raise the new note by a fifth again, giving a note with frequency $\frac{9f}{4}$. Because $\frac{9}{4}>2$, You bring it \textit{down} by an octave, or dividing by 2 which yields a note of frequency $\frac{9f}{8}$, and so forth, until you have 12 notes spanning the octave. To get the notes in the octave above, multiply all your frequencies by 2. To get the ones on the octave below, you divide by 2. However, what happens when you increase by 7 octaves? You get a note with frequency $2^{7} f=128 f$. You ought to be able to get the \textit{same} note by starting with your root note $f$ and raising it by fifths 12 times. Unfortunately, $\left(\frac{3}{2}\right)^{12}=129.74$. This small difference is known as the Pythagorean comma and it creates a serious intonation problem. As instruments became more sophisticated, new forms of intonation were sought. Still enamored by the whole number ratios of Pythagoras, the \textit{Just Intonation} used approximations to the Pythagorean ratios that used smaller integers, but this still did not fully solve the issue. For example, the Pythagorean ratios that define a major scale are \cite{bib20}:

$$\textrm{Pythagorean Intervals, Major Scale} = 
\left(1,
\frac{9}{8},
\frac{81}{64},
\frac{4}{3},
\frac{3}{2},
 \frac{27}{16},
\frac{243}{128},
2\right)$$

which, in spite of having many consonant intervals gives rise to the Pythagorean comma and the impossibility of polyphony. 
If we now try to build a scale that maximizes the number of consonant intervals having \textit{exact} rational ratios, the result is what is known as ``Just Temperament''. In this case, the major scale is defined by the following ratios:

$$\textrm{Just Intervals, Major Scale} = 
\left(1,
\frac{9}{8},
\frac{5}{4},
\frac{4}{3},
\frac{3}{2},
 \frac{10}{6},
\frac{15}{8},
2\right)$$
On the surface, this intonation is just as harmonious as the Pythagorean (same perfect 4th, 5th and octave). It also uses only rational numbers, but they use smaller integers: $\frac{81}{64}\rightarrow\frac{5}{4}$, $\frac{27}{16}\rightarrow\frac{10}{6}$ and $\frac{243}{128}\rightarrow\frac{15}{8}$. However, let's compare the scales of C and D major in just intonation:

\begin{table}[h!]
\begin{tabular}{llllllll}
 C&D  &E  & F &G &A &B &C  \\ 
1 & $\frac{9}{8}$ & $\frac{5}{4}$  &$\frac{4}{3}$  &$\frac{3}{2}$ &$\frac{10}{6}$&$\frac{243}{128}$ & 2 \\
D&E  &F\#  & G & A& B&C\# & D\\ 
$\frac{9}{8}$ &$\frac{81}{64}$ &$\frac{45}{32}$  & $\frac{3}{2}$ & $\frac{27}{16}$& $\frac{15}{8}$&$\frac{135}{64}$& $\frac{9}{4}$\\
\end{tabular}
\end{table}
The problem is clear: we can see that the notes E, A and B don't have the same frequencies! This means it would be impossible for transposing instruments to play simultaneously. This persistent difficulty was eventually solved by a brute force compromise: The size of every interval in the 12 note octave was forced to be equal. This solution, known as 12 tone equal temperament (TET) was developed independently by Zhu Zaiyu (1584) \cite{bib5} and Simon Stevin (1585) \cite{bib6} and lies at the heart of modern Western Harmony. In this intonation, only the octaves remain exact, and every other interval is slightly flat or sharp, as we can see by comparing a scale of C Major in Just Intonation and in 12 TET:

\begin{table}[h!]
\begin{tabular}{lllllllll}
Just Intonation &C&D  &E  & F &G &A &B &C  \\ 
&1 &$\frac{9}{8}$ & $\frac{5}{4}$  &$\frac{4}{3}$  &$\frac{3}{2}$ &$\frac{10}{6}$&$\frac{15}{8}$ & 2 \\
Equal Temperament&C&D  &E  & F & G& A&B & C\\ 
& 1&$2^{2/12}$ &$2^{4/12}$  & $2^{5/12}$ &$2^{7/12}$&$2^{9/12}$&$2^{11/12}$& $2^{12/12}$\\
$\%$ Difference&0&-0.23&0.79&0.11&-0.11&0.91&0.68&0\\
\end{tabular}
\end{table}
This 12 TET system is what allows musicians to modulate (change keys) freely and polyphony - as we know it in the western culture - to exist and flourish, even if it comes at a loss of the consonant frequency ratios. The percentage differences may seem small and fussing about them could therefore be seen as pedantic, but they are significant when we consider timbre; every harmonic present in a note will also be shifted, and this can cause beats to become more or less prevalent in the perception of sound when chords are played. It is worthy of note that while eaqual temperament was developed in 1584, it was not really widely adopted until the 19th century, where the last bastions of resistance among composers finally yielded. For a detailed discussion of the contrast between just intonation and equal temperament using the music of Chopin (composed using instruments tuned in Just Temperament), see \cite{bib21}.

There is a central question that remains: why 12 notes to span the octave? Why not 13, 19 or 36, say? The answer is that it doesn't \textit{have} to be 12 notes to the octave at all! As we discuss in section 2, when we formulate harmony as an ordered phase that arises from the interaction of sounds, all manner of octave divisions arise. The key distinction rests on how we define dissonance, and that will vary from culture to culture.  This is a tantalizing insight: when students learn ``music theory'' they are essentially learning the harmonic style of 17th century European composers. Thinking of harmony in the same way we think about ordered phases in nature brings all intonations observed across cultures to sit at the same table.

 The outline of this paper is as follows: In section 1 we will delve into the mathematical formulation of dissonance, which leads to the concept of ``roughness'' and discuss the importance of timbre. In section 2, we investigate the claims made by Berezovsky \cite{bib1} that it is possible to obtain harmony via the interplay of a musical internal energy (our dissonance function) and entropy, mediated by a constant parameter that is analogous to temperature. We give new analytical arguments, confirmed by numerical computation, to explain why the ordered phase has the number of notes that it does, and clearly show that some of the conditions Berezovsky identified as necessary for this formulation to work in fact are not, meaning the model is more robust than previously stated. We also extend his choice of timbre (Sawtooth with no further comment) to generalize the method. In section 3 we present our new results and analyze Indonesian intonation using the thermodynamic analogy. 

\section{Timbre and dissonance}
Whenever we hear a ``tone'' produced by an acoustic instrument we are never only hearing a single frequency. Each note consists of a fundamental frequency $f_{1}$ of amplitude $A_{1}$, and a number of harmonics of that fundamental, each with an amplitude $A_{n}=a(n)A_{1}$ and a frequency $f_{n}=\varphi(n)f_{1}$ with $n>1$. The details of both $a(n)$ and $\varphi(n)$ are what specify the \textit{timbre} of a given note. Timbre explains why the ``same'' note (meaning a note of a given fundamental frequency $f_{1}$) sounds different when played by different instruments. In other words, each instrument has its own unique timbre.
\begin{figure}[!h]
\caption{{\bf Superposition of two pure sinusoids.}
\includegraphics[width=2in]{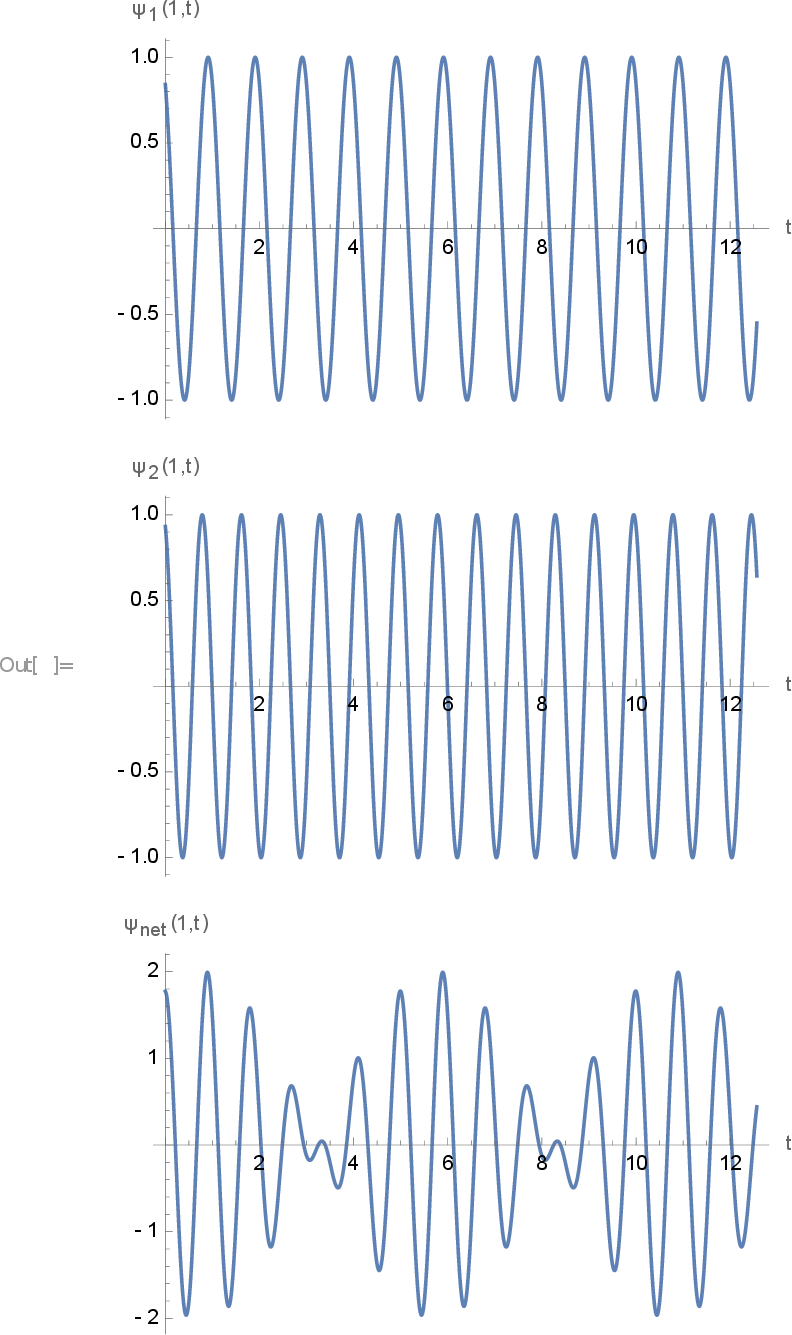}
This is a time plot of two plane waves  $\psi_{1}(x,t)$ and $\psi_{2}(x,t)$ of different frequency and wavelength but equal amplitude, as well as their superposition $\psi_{net}(x,t)=\psi_{1}(x,t)+\psi_{2}(x,t)$ taken at the point $x=1$ in arbitrary units.}
\label{figoverlap}
\end{figure}

We can think of music as a collection of notes or tones that are played sequentially and purposefully in time. How this music is perceived will depend on how a listener perceives the interactions between these notes. In this context, it becomes useful to define the \textit{dissonance} between notes. Let's begin by considering only two, pure sinusoidal oscillations. These can only be obtained from an electronic instrument. For those two sounds we can say that ``dissonance'' quantifies the extent to which they will sound unpleasant or ``rough'' when played together. Indeed, Helmholtz attempted to define dissonance entirely based on the beat frequency that results from overlapping two pure sinusoidal waves \cite{bib11}. For those not familiar, a quick review is in order: consider two waves, $\psi_{1}(x,t)$ and $\psi_{2}(x,t)$, moving to the right. The waves have equal amplitude $A$, but different wavelengths ($\lambda_{1}\neq\lambda_{2}$) and frequencies ($f_{1}\neq f_{2}$):
\begin{eqnarray}
\psi_{1}(x,t) & = & A\sin\left(k_{1}x-\omega_{1}t\right) ,\\
\psi_{2}(x,t) & = & A\sin\left(k_{2}x-\omega_{2}t\right) ,
\end{eqnarray}
where, for writing economy we have introduced the wave number $k\equiv\frac{2\pi}{\lambda}$ and the angular frequency $\omega\equiv2\pi f$. When these two waves overlap, the principle of superposition tells us that the resulting wave $\psi_{net}(x,t)=\psi_{1}(x,t)+\psi_{2}(x,t)$. Using the well known angle difference identity $\sin(\alpha-\beta)=\sin\alpha\cos\beta-\cos\alpha\sin\beta$, it is straightforward to show that
\begin{equation}
\label{psinet}
\psi_{net}(x,t)=2A\cos\left[\frac{x}{2}(k_{1}-k_{2})-\frac{t}{2}(\omega_{1}-\omega_{2})\right]\sin\left[\frac{x}{2}(k_{1}+k_{2})-\frac{t}{2}(\omega_{1}+\omega_{2})\right].
\end{equation}
Equation (\ref{psinet}) is illustrated in Fig. (\ref{figoverlap}) and the bottom plot corresponds to what a microphone would capture if two overlapping pure sinusoids were recorded. We can see from Equation (\ref{psinet}) there are two basic frequencies that characterize $\psi_{net}(x,t)$: First, we can see a fast oscillation at a frequency $F=\frac{f_{1}+f_{2}}{2}$ coming from the sine term, which would be the frequency captured by a microphone. It is important to note that, because the ear functions as a frequency spectrum analyzer, $F$ is \textit{not} necessarily the frequency a human being will hear. $F$ will only be be the frequency heard provided that $|f_{1}-f_{2}|$ is smaller than the resolution of the cochlea. When that happens, a ``fused'' tone $F$ will be heard. However, as $|f_{1}-f_{2}|$ increases above cochlear resolution, $f_{1}$ and $f_{2}$ can be interpreted by the ear. This is what allows a listener to hear the multiple notes that form a chord. For greater insight into how perception goes beyond superposition, see \cite{bib23}.

In addition, Fig. (\ref{figoverlap}) shows there is a slower variation in amplitude coming from the cosine term at a frequency  $f=\frac{f_{1}-f_{2}}{2}$. Now, because every cycle of the amplitude envelope contains a crest and a trough, and they both represent a maximum in amplitude (and thus intensity) of the wave, the loudness of the wave will pulse at twice the value of $F$. Indeed, $2F$ is known as the beat frequency, $f_{B}$, between two pure tones of frequencies $f_{1}$ and $f_{2}$, defined as
\begin{equation}
\label{beats}
f_{B}=|f_{1}-f_{2}|,
\end{equation}
where the absolute value ensures we are only considering the magnitude of the difference in frequencies. If the beat frequency is not a whole number multiple of either one of the frequencies being overlapped, and yet it is large enough to be perceived as an audible tone, this beat frequency will be perceived as ``dissonance''. It should be noted that this is not exactly the same thing as musical dissonance, which is more nuanced and has a more complex meaning and use in music. The Helmholtz definition refers to \textit{roughness}; a phenomenon due to the beats that occur when the two sounds are played together. Based on this definition, one would expect that when the two notes are of the same frequency (unison) roughness is zero. As the difference in frequency between the notes increases, so does roughness,  peaking not far past unison and then decreasing to zero as the difference in frequency between the two pure sinusoids increases.

The perception of dissonance between two real notes is more nuanced and depends on a number of factors:
\begin{enumerate}
  \item The physical superposition between the pure harmonics that comprise the notes - their spectral interference and how the notes beat against each other - and thus the amount of roughness in the Helmholtz sense.
  \item Even then, surprises may arise because beat frequencies may give rise to the perception of sound with little to no roughness. For example, if you play two pure sinusoidal frequencies of $f_{1}=220$ Hz ($A_{3}$ as it is referred on a piano) and $f_{2}=330$ Hz (a minutely slightly sharp version of $E_{4}$, which in modern, 12-tone equal temperament corresponds to a frequency of 329.6 Hz ), the beat frequency will be $f_{B}=|f_{1}-f_{2}|=110$ Hz. This is exactly half the frequency of $f_{1}$. In music, we would say that $f_{B}$ is exactly one \textit{octave} below $f_{1}$, and would create the perception of a bass note of $A_{2}$. This phenomenon is also known as the ``missing fundamental''.
  \item The notes produced by acoustic instruments are never pure sinusoidal oscillations; each one is accompanied by a harmonic series of overtones of different amplitudes which constitute the \textit{timbre} of the note, and the perception of dissonance when they are played together will depend on the specific timbre of the notes. If we were to play an $A_{3}$ and an $E_{4}$ on a guitar tuned to modern 12 tone equal temperament, there would be multiple harmonics beating against each other - including a very good approximation to the missing fundamental at 109.6 Hz - and not all of the resulting beat frequencies would create a sensation of "roughness".
  \item Last and by no means least, the perception of musical dissonance will depend significantly on the upbringing, culture and listening habits of the individual.
\end{enumerate}
We will have to consider the effect of cultural conditioning a bit later; for now, we will only consider 1. above, and use the resulting perception of roughness among Western listeners as our measure of dissonance, in accord with Plompt and Levelt \cite{bib2}.

If we consider two notes with fundamental frequencies $f_{i}$ and $f_{j}$, we can define a strength of dissonance interaction between them that we call $D_{ij}$, and can then write the total dissonance $D_{total}$ as
\begin{equation}
\label{dtot}
D_{tot}=\sum_{i,j}D_{ij},
\end{equation}
where we will be taking the sum over all the tones in the music being considered. In order to calculate this total dissonance we must first find a way to calculate the various $D_{ij}$ coefficients. This, it turns out, can be very subtle and will depend crucially on cultural factors that determine whether or not two notes played together are perceived as dissonant or not.

\begin{figure}[!h]
\caption{{\bf Dissonance between two real notes with a sawtooth timbre.}
\includegraphics[width=3in]{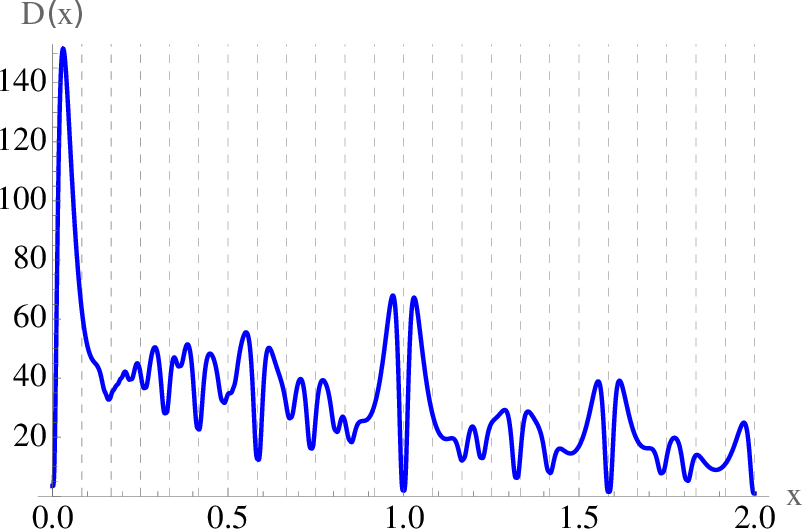}
Plot of the dissonance function $D(x)$ as a function of the pitch difference $x$. In this case, $w_{c}=0.03$, and we see that $D$ is zero for the unison, and zero again at the octaves ($ x=1$ and $x=2$). Vertical dashed lines represent the twelve-tone equal temperament (12 TET) octave division.}
\label{fig1}
\end{figure}

We can begin by seeking a function $D(f_{i},f_{j})$ that quantifies dissonance between two notes of fundamental frequencies $f_{i}$ and $f_{j}$. It will be important to know the timbre of these notes, and how strong the interaction of these notes is perceived. If we define a dissonance interaction coefficient $\alpha_{ij}$, we can write
\begin{equation}
\label{dij}
D_{ij}=\alpha_{ij}D(f_{i},f_{j})
\end{equation}
where, if two notes overlap in time, the corresponding value of $\alpha_{ij}$ would be higher than if they don't.  In the literature one finds different ways in which we can compute $D(f_{i},f_{j})$. Following \cite{bib4}, \cite{bib12} we first calculate the \textit{pure tone} roughness between two pure frequencies $f_{k}$ and $f_{l}$ of amplitudes $A=1$ that have no harmonics, i.e. $a_{n}=0$ for $n>1$; these two frequencies would have to be generated electronically. We define this roughness function for pure tones as $d(f_{k},f_{l})$,
\begin{equation}
\label{putetoned}
d(f_{k},f_{l})=d\left(\Delta x\right),
\end{equation}
where we define the \textit{pitch difference}, $\Delta x$, as
\begin{equation}
\label{pitchd}
\Delta x=\log_{2}\left(\frac{f_{k}}{f_{l}}\right).
\end{equation}
This definition of $\Delta x$ is convenient to identify octaves; in music, an octave is a span of frequencies between $f$ and $2f$ that are perceived as being the ``same'' note because every harmonic of $2f$ is contained in the harmonics of $f$. Here, although we are dealing with pure tones that have no harmonics we can still count octaves conveniently: if $f_{k}=2f_{l}$, we have that $\Delta x =1$.  We can see in the literature that the pure tone roughness function will increase as a function of $\Delta x$ up to a critical value $\Delta x=w_{c}$, after which it will decrease for increasing values of $\Delta x$. It is essential that the two-tone roughness satisfies the condition $d(\Delta x=1)=0$. Otherwise, the implication is that roughness will be perceived for a perfect octave, which is unphysical given that roughness is a consequence of beats. There are many different ways in which authors parametrize the two-tone roughness function in the literature, and it pays to be careful. Plomp and Levelt \cite{bib2} correctly plot a two-tone \textit{consonance} function which is zero at both $\Delta x =0$ and $\Delta x=1$.  They do not give an analytical expression for their function but in the literature several analytical expressions have been fit to the qualitative expectations found in the literature \cite{bib2}. The following expression for the two-tone roughness was used as a fit in \cite{bib1}:
\begin{equation}
\label{dberezov}
d(\Delta x)=\frac{1}{w_{c}}\exp\left[-\ln\left(\frac{\Delta x}{w_{c}}\right)\right]^{2}.
\end{equation}
Our aim is to use Eq(\ref{dberezov}) to calculate the dissonance between two real notes, taking into account their timbres. However, to do so we need to have the value of $w_{c}$, and there is a subtlety in \cite{bib1} worth clarifying. Human hearing ranges from a minimum audible frequency $f_{min}\sim20$ Hz to a maximum audible frequency $f_{max}\sim20,000$ Hz. In \cite{bib1}, it is reported that for frequencies in the range of $f_{min}$, Eq (\ref{wcbere}) yields $w_{c}\sim0.5$, while for the higher audible frequencies $w_{c}\sim0.002$. It is then proposed that 
\begin{equation}
\label{wcbere}
w_{c}=0.67\min(f_{k},f_{l})^{-0.68}.
\end{equation}
However, in the range from 10,000 to 20,000 Hz, Eq (\ref{wcbere}) yields $w_{c}\sim0.0008$. We believe this is a typo, and the parametrization for $w_{c}$ that accurately gives the range of observed values reported in \cite{bib1} is given by
\begin{equation}
\label{wcus}
w_{c}=5.5\min(f_{k},f_{l})^{-0.68}.
\end{equation}
We remark on this discrepancy because as we will see, how we quantify dissonance plays a crucial role in the analogy we draw between thermodynamics and harmony. We can now finally calculate the dissonance between two real notes by summing over all pairs of pure partials:
\begin{equation}
\label{dpair}
D(f_{i},f_{j})=\sum_{k,l}I_{kl}d(\phi(k)f_{i},\phi(l)f_{j}),
\end{equation}
where the $I_{kl}$ is a parameter given by
\begin{equation}
\label{loud}
I_{kl}=\min(a(k),a(l))^{0.606}
\end{equation}
that estimates the perceived ``loudness'' of the pair of harmonics \cite{bib4}.  It is useful to recall that a given note of frequency $f_{i}$ will have many audible partials, characterized by parameters of amplitude $a(n)$ and frequency $\phi(n)$ which together specify the note's timbre. As an example, we can use Eq (\ref{dpair}) to calculate the dissonance between two real notes. We assume the timbre of the notes is that of a sawtooth waveform, with $a(n)=1/n$, and add up to 10 partials to obtain $D(x)$ as shown in Fig~\ref{fig1}.

Other minima are clearly visible in Fig~\ref{fig1}. For example, the second largest minimum occurs at $x=0.585$. This means that $f_{a}=3f_{b}/2$, which corresponds to a \textit{perfect fifth} interval, which is commonly used in music. Other minima occur at other common music intervals, for example when $ x=0.322$ which corresponds to a \textit{major third} where $ f_{a}=5f_{b}/4$, or when $ x=0.415$ which happens when $ f_{a}=4f_{b}/3$ and corresponds to a \textit{perfect fourth}, etc.

One interesting result that can be obtained from Fig~\ref{fig1} is that for the sawtooth timbre, there are a total of 12 minima within an octave, each one corresponding to pitches with simple, rational frequency ratios ($2/1, 3/2, 4/3, 5/4\cdots$). We can use the minima in the dissonance function to understand why the octave is divided into 12 pitches in Western music. Other cultures have a different concept of dissonance, and therefore as we will see later on, the minima distribution of the corresponding $D(x)$ is different, leading to octaves that can be divided in 5, 7 and other numbers of pitches.

\begin{figure}[!h]
\caption{{\bf Roughness between two pure sinusoidal tones.}
\includegraphics[width=4in]{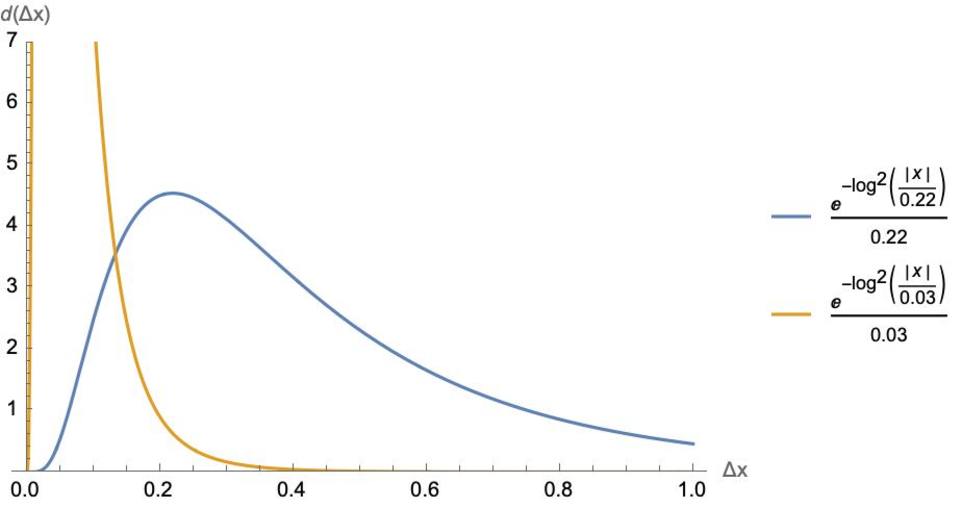}
Plot of the two pure tone roughness function $d(\Delta x)$ as a function of the pitch difference $\Delta x$ for $w_{c}=0.03$ and  $w_{c}=0.22$, where we see that as the pitch difference increases, the two pure tone roughness decreases to zero only for the smaller value of $w_{c}$.}
\label{fig2}
\end{figure}

One final remark is in order: Eq (\ref{dberezov}) satisfies the necessary condition $d(\Delta x =0) = 0$. However, it does not satisfy $d(\Delta x =1)=0$. For small values of $w_{c}$, $d(\Delta x=1)\approx 0$, but as the value of $w_{c}$ increases this approximation may no longer valid. For example, if $w_{c}$ is small, say $w_{c}=0.03$, Eq (\ref{dberezov}) satisfies the condition that $d(\Delta x=1)\approx0$. However, if we increase to $w_{c}=0.22$, the condition is no longer satisfied, as shown in Fig~\ref{fig2}. This deviation from zero at $\Delta x=1$ is something worth investigating to make sure it bears no effect on the results presented in \cite{bib1}. To do so, we can multiply Eq(\ref{dberezov}) by an arbitrary factor $q(x)$ chosen to force the zero at $\Delta x =1$:
\begin{equation}
\label{qfunc}
q(x)=\frac{1}{1+e^{100(x-0.8)}}.
\end{equation}

\begin{figure}[!h]
\caption{{\bf Roughness between two pure sinusoidal tones.}
\includegraphics[width=3in]{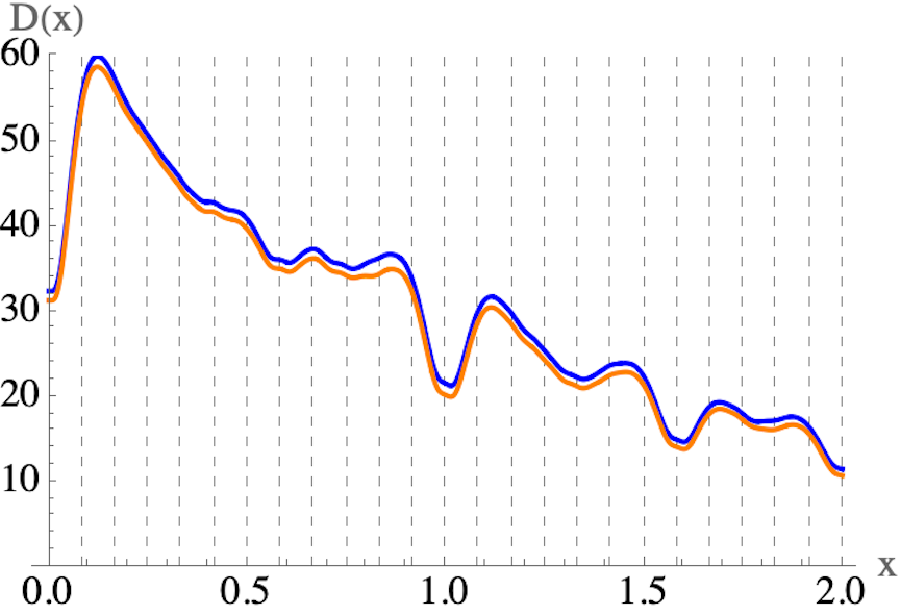}
Plot of D(x) for both $d(\Delta x=1)\approx 0$  (blue line) and $d(\Delta x=1)=0$ (orange line).} 
\label{fig3}
\end{figure}

 If we use Eq(\ref{dpair}) to calculate $D(x)$ for the pure two-tone dissonance forced to be equal to zero at $x=1$ and compare top the result obtained without this condition being met i.e. using Eq(\ref{dberezov}), shown in Fig~\ref{fig2}, we obtain Fig~\ref{fig3}, from which we can see that there is no inherent problem with the fact that Eq(\ref{dberezov}) is not exactly equal to zero at $x=1$.
 \pagebreak

\section{Music as a thermodynamic system}
Thermodynamics is a powerful tool that allows us to understand how order arises from disordered states of matter via the minimization of the Helmholtz free energy $F$ given by
\begin{equation}
\label{helmfree}
F=U-TS,
\end{equation}
where $U$ is the internal energy of the thermodynamic system, and $TS$ is the amount of that energy that is disordered, where $S$ is what we call the \textit{entropy} and $T$ denotes the thermodynamic temperature, which is a parameter that mediates the tradeoff between the decreasing $U$ and the increasing $S$ which yields a minimum value for $F$.
The tantalizing idea explored by Berezovsky was that it might to be possible to quantify harmony in a similar way: ordered phases of sound arising from disordered sound. In this case, the key point was to introduce a musical entropy calculated in terms of \textit{information} - the number of notes per octave, and to allow it to grow against a decreasing musical energy he identified as the total dissonance, calculated using Eqs (\ref{dij}) and (\ref{dpair}). In this calculation, we are therefore minimizing the value of a ``musical'' free energy $F_{M}$ given by
\begin{equation}
\label{helm}
F_{M}=D_{tot}-TS,
\end{equation}
where $T$ is a parameter we call temperature by analogy with thermodynamics, which in this case mediates the tradeoff between $D_{tot}$ and $S$.
\subsection*{Calculation of the probability distribution function of relative pitches.}
The idea is to calculate $P(x)$ - the distribution function of  relative pitches $x$ that occur in music - such that the free energy given by Eq (\ref{helm}) is minimized, subject to the constraint that $P(x)$ is normalized. The relative pitches $x$ in music, are defined as
\begin{equation}
\label{xdef}
x=\log_{2}\left(\frac{f}{f_{ref}}\right).
\end{equation}
We calculate the total dissonance $D_{tot}$ as
\begin{equation}
\label{totaldiss}
D_{tot}=\frac{1}{2}\int_{0}^{1}dy \int_{0}^{1} P(x)D_{P}(x-y)P(y)dx,
\end{equation}
where $D_{p}$ is for now defined by adding over all octaves as:
\begin{equation}
\label{dp}
D_{p}(x)=\sum_{n=-\infty}^{\infty}D(x+n),
\end{equation}
We will explore altering this definition later.
\begin{figure}[!h]
\caption{{\bf Total Dissonance between two real notes with a sawtooth timbre.}
\includegraphics[width=3in]{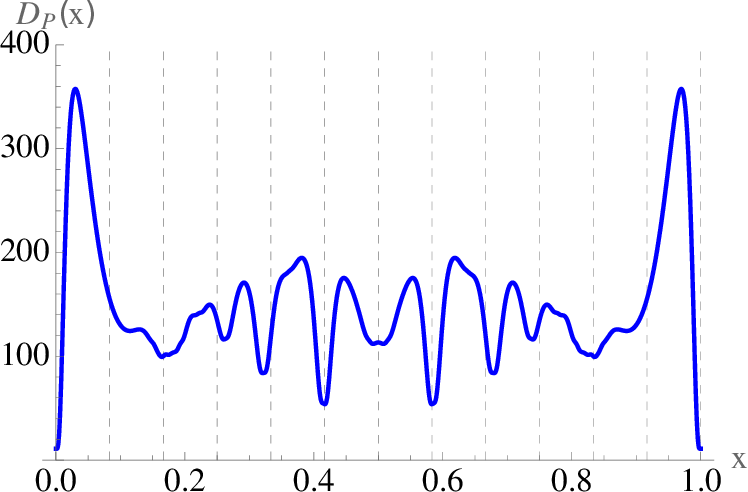}
Plot of the total dissonance function $D_{P}(x)$ as a function of the pitch difference $x$. In this case, $w_{c}=0.03$, and we see that $D$ is zero for the unison, and zero again at the octaves ($ x=1$ and $x=2$). We also note the symmetry about the point $x=0.5$}
\label{fig4}
\end{figure}
Eq (\ref{dp}) is illustrated in Fig. \ref{fig4}. Similarly, the entropy is written as:
\begin{equation}
\label{ent}
S=-\int_{0}^{1}P(x)\ln P(x)dx.
\end{equation}
The functional we must minimize then becomes
\begin{equation}
\label{ }
F=\frac{1}{2}\int_{0}^{1}dy\int_{0}^{1}P(x)D_{P}(x-y)P(y)dx+T\int_{0}^{1}P(x)\ln P(x)dx,
\end{equation}
subject to the normalization condition
\begin{equation}
\label{normal}
N=\int_{0}^{1}P(x)dx-1=0.
\end{equation}
Using a Lagrange multiplier $\mu$, we want to calculate an extremum of a new functional $F^{*}$ given by
\begin{eqnarray}
\label{lagrange}
F^{*} & = &D_{tot}-TS+\mu N\nonumber \\
 & = & \int_{0}^{1}dy\int_{0}^{1} f^{*}(x,P)dx.
\end{eqnarray}
This gives
\begin{eqnarray}
\frac{d}{dx}\left(\frac{\partial f^{*}}{\partial P^{\prime}}\right) & = & \frac{\partial f^{*}}{\partial P}\nonumber \\
\label{euler}
\Rightarrow \frac{\partial f^{*}}{\partial P} & = & 0,
\end{eqnarray}
where
\begin{equation}
\label{fstar}
f^{*}=\left[\int_{0}^{1}P(x)D_{P}(x-y)P(y)dy\right]+TP(x)\ln P(x)-\mu\left[P(x)-1\right].
\end{equation}
putting Eq (\ref{fstar}) into (\ref{euler}), we obtain
\begin{eqnarray}
\frac{\partial f^{*}}{\partial P} & = &\int_{0}^{1}D_{P}(x-y)P(y)dy+T\ln P(x)+T- \mu\nonumber \\
 & = & 0
\end{eqnarray}
setting the Lagrange multiplier $\mu =T$ we then obtain that
\begin{eqnarray}
\label{almost}
\ln\left[\frac{P}{P_{0}}\right]&=&-\frac{1}{T}\int_{0}^{1}D_{P}(x-y)P(y)dy\nonumber\\
\Rightarrow\frac{P}{P_{0}}&=&\exp\left[-\frac{1}{T}\int_{0}^{1}D_{P}(x-y)dy\right],
\end{eqnarray}
where $P_{0}$ is a normalization constant given by
\begin{equation}
\label{ }
P_{0}=\int_{0}^{1}dz \exp\left[-\frac{1}{T}\int_{0}^{1}dy D_{p}(z-y)P(y)\right],
\end{equation}
and therefore, the equilibrium distribution of relative pitches that minimizes the musical free energy, $P(x)$, is given by
\begin{equation}
\label{meanfield}
P(x)=\frac{\exp\left[-\frac{1}{T}\int_{0}^{1} D_{p}(x-y)P(y)dy\right]}{\int_{0}^{1}dz \exp\left[-\frac{1}{T}\int_{0}^{1} D_{p}(z-y)P(y)dy\right]},
\end{equation}

Equation (\ref{dp}) ensures that a given musical system is preserved across octaves so we can define $P(x+n)=P(x)$, where $n$ is an integer, i.e. a melody will be recognized as the same when played in any octave. Now, because $D_{p}(x)$ is an even function of $x$, it follows that its Fourier expansion may be written as
\begin{equation}
\label{dpfourier}
D_{p}(x)=\sum_{n=0}^{\infty}d_{n}\cos(2 n \pi x).
\end{equation}
 If we insert Eqn (\ref{dpfourier}) into (\ref{meanfield}), we have:
\begin{eqnarray}
P(x)& = & \frac{1}{P_{0}}\exp\left[-\frac{1}{T}\int_{0}^{1} D_{p}(x-y)P(y)dy\right]\nonumber \\
 & = &  \frac{1}{P_{0}}\exp\left[-\frac{1}{T}\int_{0}^{1}\sum_{n=0}^{\infty}d_{n}\cos(2 n \pi (x-y)P(y)dy\right]\nonumber\\
 &=&\frac{1}{P_{0}}\exp\left[-\frac{1}{T}\sum_{n=0}^{\infty}d_{n}\int_{0}^{1}\left(\cos(2\pi nx)\cos(2\pi n y)+\sin(2\pi n x)\sin(2\pi n y)\right)P(y)dy\right]\nonumber\\
 %&=&\frac{1}{N}\exp\left[-\sum_{n=0}^{\infty}d_{n}\left(\frac{\cos(2\pi n x)}{T}\int_{0}^{1}dy\cos(2\pi ny) P(y)+\frac{\sin(2\pi n x)}{T}\int_{0}^{1}dy\sin(2\pi ny) P(y)\right)\right]\nonumber\\
 \label{pofx}
 &=&\frac{1}{P_{0}}\exp\left[-\sum_{n=0}^{\infty}d_{n}\left(\frac{\alpha_{n}\cos(2\pi n x)}{T}+\frac{\beta_{n}\sin(2\pi n x)}{T}\right)\right],
\end{eqnarray}
where
\begin{equation}
\label{alfa}
\alpha_{n}=\int_{0}^{1}\cos(2\pi ny) P(y)dy,
\end{equation}
and
\begin{equation}
\label{beta}
\beta_{n}=\int_{0}^{1}\sin(2\pi ny) P(y)dy.
\end{equation}

% For figure citations, please use "Fig" instead of "Figure".

% Place figure captions after the first paragraph in which they are cited.
%\begin{figure}[!h]
%\caption{{\bf Bold the figure title.}
%Figure caption text here, please use this space for the figure panel descriptions instead of using %subfigure commands. A: Lorem ipsum dolor sit amet. B: Consectetur adipiscing elit.}
%\label{fig1}
%\end{figure}
%\begin{figure}
%\begin{center}
%\includegraphics[width=2in]{file_path}
%\caption{ }
%\label{ }
%\end{center}
%\end{figure}

% Results and Discussion can be combined.
\section*{Results}
% Place tables after the first paragraph in which they are cited.
\subsection*{Periodicity Prediction for $P(x)$}
Both of the integrals given in Eqs (\ref{alfa}) and (\ref{beta}) are seen to be approximately constant and of $ O(1)$ over a wide range of values of $n$. The presence of an audible upper limit of $\approx20,000$ Hz implies that in practice, the sum in Eq (\ref{pofx}) will not have an infinite number of terms. Intuitively, one would expect that the behavior of the entire $P(x)$ will be dominated by the terms in the sum with the largest amplitudes. Given the exponential function in Eq (\ref{meanfield}), the largest term present in Eq (\ref{dpfourier})will significantly dominate over all others. Indeed, at sufficiently high temperatures, all terms in the sum (Eq(\ref{pofx})) will be much smaller than 1, and $P(x)$ will tend to a constant. As the temperature decreases, the largest term in the sum, $n_{max}$ set by $D_{P}(x)$ will cross the unity threshold and dominate over all other terms, leading to an $n_{max}$-periodic $P(x)$ solution. 
%\end{itemize}
It is then expected that below a certain critical temperature, $T_{c2}$, to very good approximation over the entire range $0\leqslant x\leqslant1$, we can write
\begin{equation}
\label{papprox}
P(x)\simeq\exp\left[-\frac{d_{max}}{T}\sin(2\pi n_{max} x)-\frac{d_{max}}{T}\cos(2\pi n_{max} x)\right],
\end{equation}
where $d_{max}$ is the maximum Fourier coefficient of $D_{p}$ and $n_{max}$ tells us for which harmonic this maximum occurs. Above $T_{c2}$, $P(x)$ will tend to a constant.

\begin{figure}[!h]
\caption{{\bf Fourier spectrum of $D_{P}(x)$ as a function of $k_{n}$.}
\includegraphics[width=5in]{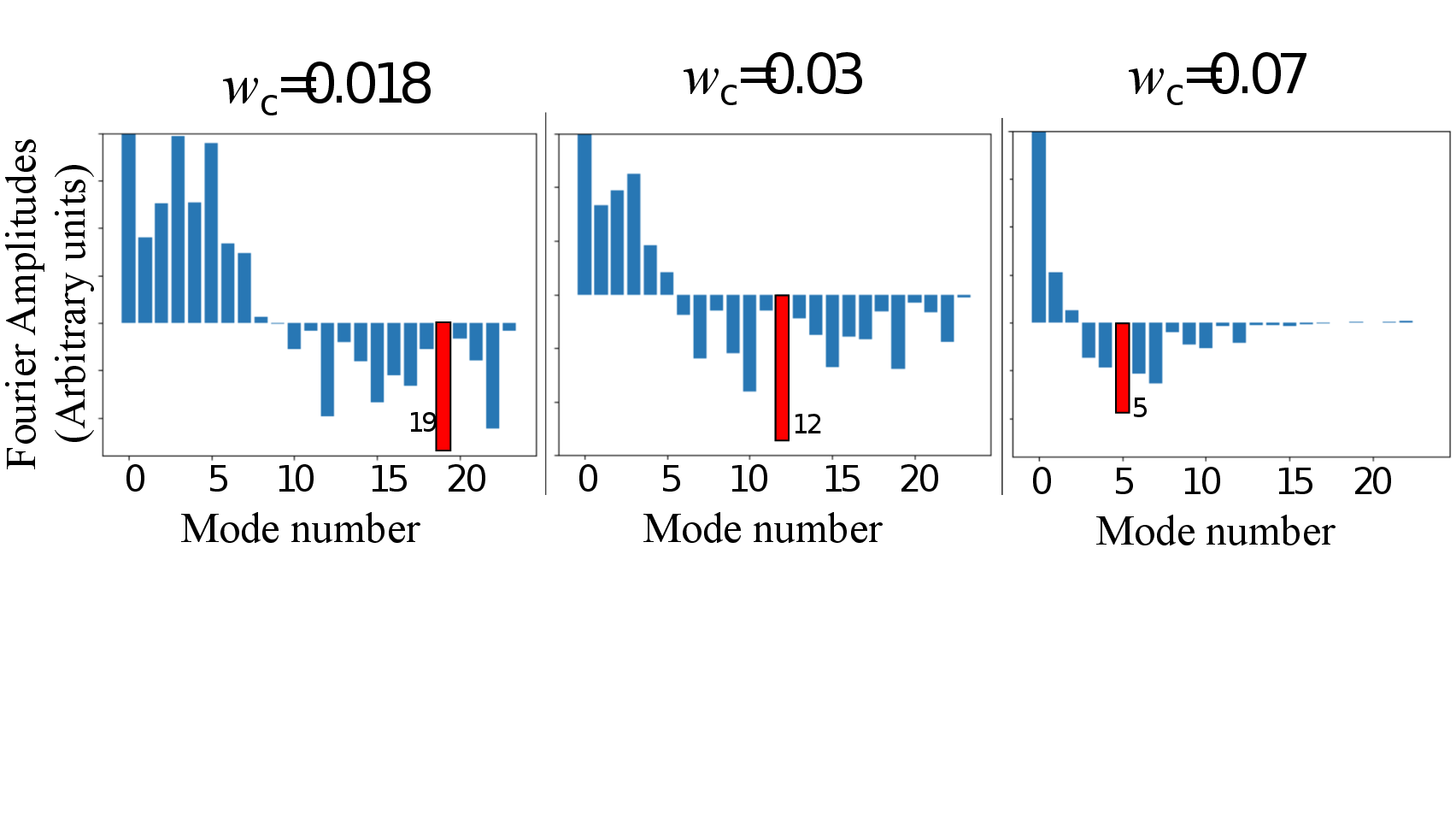}
Plot of the Fourier representation of $D_{P}(x)$ as a function of $k$ with a sawtooth timbre structure and harmonic amplitudes decreasing as $1/n$, for three values of $w_{c}$,  and we see that largest Fourier coefficient is the 19th when $w_{c}=0.018$, the 12th when $w_{c}=0.03$ and the 5th when $w_{c}=0.07$.}
\label{fig5}
\end{figure}

This simple analytical interpretation has been borne out by various numerical experiments shown in Figures.(\ref{fig5}), (\ref{fig:6}) and (\ref{fig:full_scan}). 

These plots illustrate that the number of peaks of $P(x)$ corresponds to $n_{max}$, and therefore depends on the value of $w_{c}$. Indeed, the width of the peaks of $D_{P}(x)$ is directly proportional to the value of $w_{c}$, and therefore the value of $w_{c}$ also determines how many notes will fill an octave.

As we have shown numerically, the number of peaks in the resulting $P(x)$ sensitively depends on $w_{c}$. This is clear from the analysis presented, as a small $w_{c}$ implies narrower features in both $D(x)$ and $D_{P}(x)$. Indeed, the width of the leading features near $x=0$ and $x=1$ of these functions scale directly with $w_{c}$. Narrow features in $D(x)$ and $D_{P}(x)$ imply higher frequency spectral features and hence larger dominant Fourier components in $D_{P}(x)$. The above in turn leads to $P(x)$ solutions with a higher frequency $P(x)$ periodicity. Conversely, going to larger $w_{c}$ values results in lower Fourier frequency $D_{P}(x)$ decompositions, and hence lower frequency $P(x)$ periodicity, as shown explicitly in the examples above. See reference \cite{bib1}.\\

\subsection*{Confirming Berezovsky results}
To confirm the claim that $P(x)$ will be dominated by the predicted Fourier coefficient $n_\textrm{max}$ as described above in and shown in fig \ref{fig:6}, we performed temperature scans for low, intermediate and high $w_c$ values. The results are summarized in the fig\ref{fig:6}:

\begin{figure}[h]
\begin{center}
\includegraphics[width=5.5in]{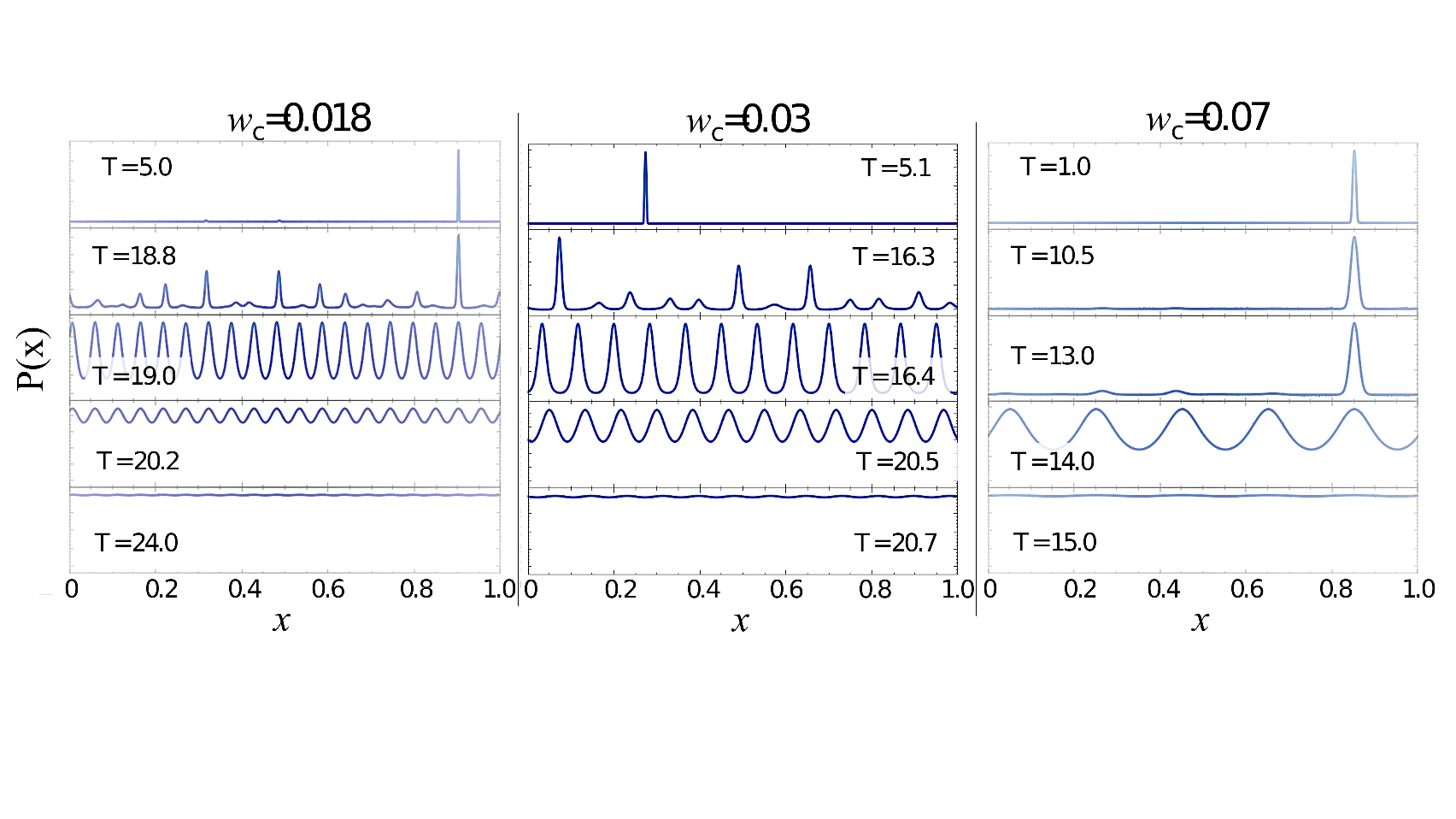}
\end{center}
\caption{$P(x)$ for selected temperatures for $w_c=0.018$,  $w_c=0.03$, and  $w_c=0.07$. Note that as temperature increases, the solution tends toward periodic with a single dominant Fourier coefficient. In the $w_c=0.018$ case,  $n_\textrm{max} = 19$. In the $w_c=0.03$ case,  $n_\textrm{max} = 12$.  In the $w_c=0.07$ case,  $n_\textrm{max} = 5$. }
\label{fig:6}
\end{figure}  
Once $P(x)$ has been numerically calculated, we performed a Fourier decomposition of $P(x)$, arriving at a series of amplitudes for the integer harmonics $k_P = 1 - 12$. Fig \ref{fig:full_scan} shows the amplitudes of these amplitudes for $w_c = 0.03$ over a range of temperatures. As also demonstrated by Berezovsky, we see the emergence of two critical temperatures, $T_\textrm{c1}$ and $T_\textrm{c2}$. For temperatures lower than $T_\textrm{c1}$, all $k_P$ contribute to $P(x)$. As the temperature increases above $T_\textrm{c1}$, we see the dominance of $n_\textrm{max} = 12$, as predicted by the analysis in Results section 1. For $T>T_\textrm{c2}$, $P(x) \approx 1$, so $k_P = 0$ (the constant coefficient, not plotted in the fig~\ref{fig:6}) is the only remaining nonzero component. 

\begin{figure}[h]
\begin{center}
\includegraphics[scale=0.4]{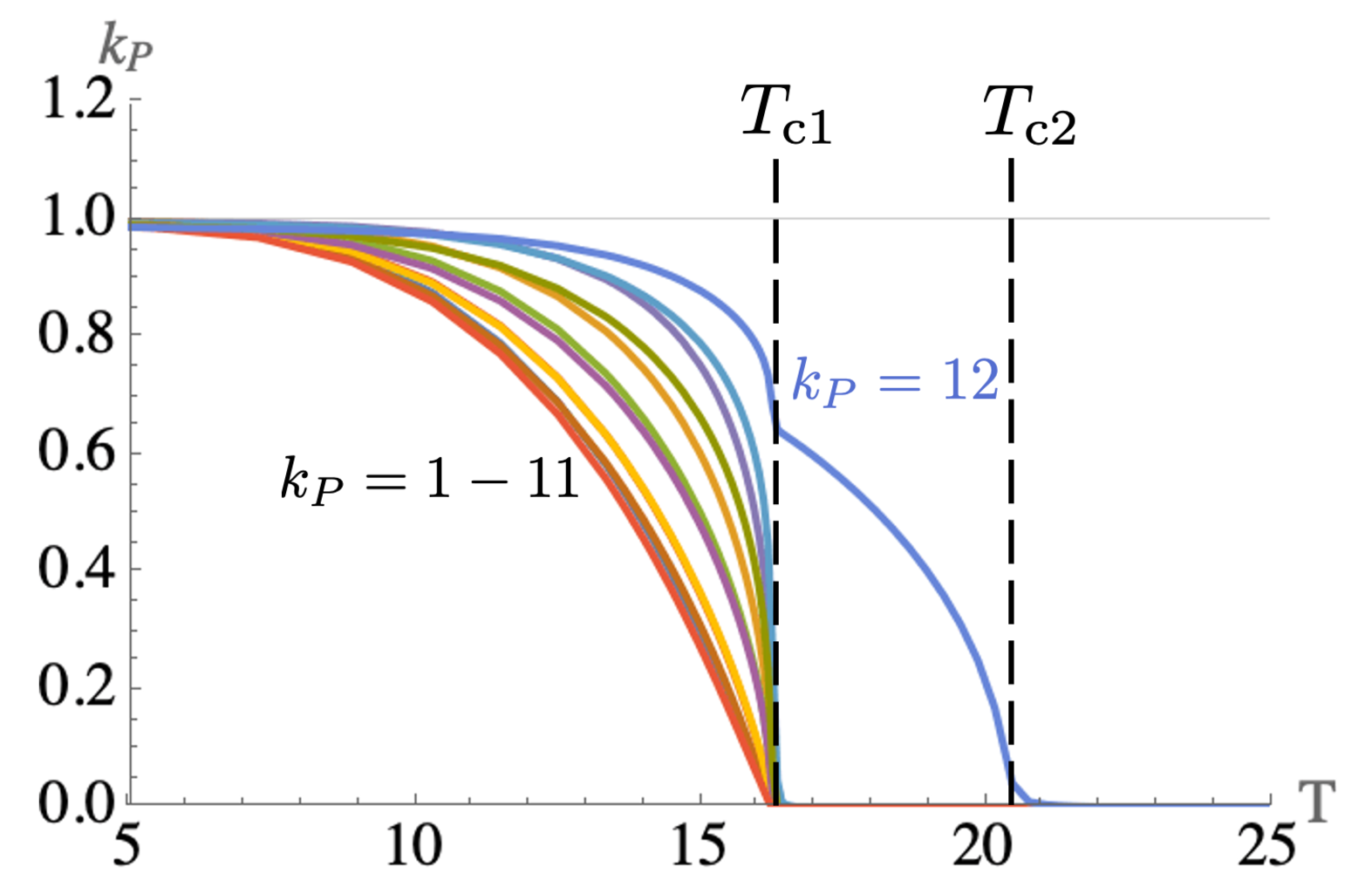}
\end{center}
\vspace*{-0.5cm}
\caption{Lowest twelve Fourier harmonic amplitudes from the decomposition of $P(x)$ over a range of temperatures using $w_c=0.03$. Between the two critical temperatures $T_\textrm{c1}$ and $T_\textrm{c2}$, the dominant component contributing to $P(x)$ is $k_P = 12$, leading to the periodic solution seen in \ref{fig:6}. }
\label{fig:full_scan}
\end{figure}  
\subsection*{Is the model robust to timber variations?}
We next investigate whether varying the timbre affects the results significantly. In addition to the sawtooth timbre already discussed, we run the same model for three alternate timbres (triangle, square and an example of human voice), keeping the periodicity over one octave, as prescribed by Eq (\ref{dp}) .

The square and triangle wave harmonic series notably contain only odd harmonics. Disregarding constant multiplicative factors, these harmonic amplitudes can be written: 

%\begin{equation}
%a_\textrm{square}(n) = \frac{1}{(2n-1)} \ \ \ \ \ \ \ \ \ \ \ \ \ \ \ \ \ \ \ \ \  a_\textrm{triangle}(n) = \frac{(-1)^n}{(2n-1)^2} 
%\end{equation}

\begin{eqnarray*}
\phi_\textrm{square}(n)  & = & \left\{ \begin{array}{cc} 1/n & \textrm{if } n \textrm{ is odd} \\
0 &  \textrm{if } n \textrm{ is even} \end{array} 
\right. 
\end{eqnarray*}

\begin{eqnarray*}
\phi_\textrm{triangle}(n)  & = & \left\{ \begin{array}{cc} (-1)^n/n^2 & \textrm{if } n \textrm{ is odd} \\
0 &  \textrm{if } n \textrm{ is even} \end{array} 
\right. 
\end{eqnarray*}

In all these tests, we include harmonics only up to $n=10$, as we did earlier with the sawtooth wave. As a final timber test, we measured the spectrum of a male human voice singing the long-``e" vowel sound and measured the first 10 harmonic amplitudes to be $$a_\textrm{voice}(n) = 
(1,
0.94,
0.63,
0.60,
0.56,
0.50,
0.60,
0.71,
1.19,
1.00)$$
taking $a(1)$ to be unity by definition. This vocal sample was sung by Roey Ben-Yoseph in the album ``A Sky Full of Ghosts" by the band Sonus Umbra where both A. Tillotson and L. Nasser performed and produced. 

\begin{figure}[h]
\begin{center}
\includegraphics[width=5.5in]{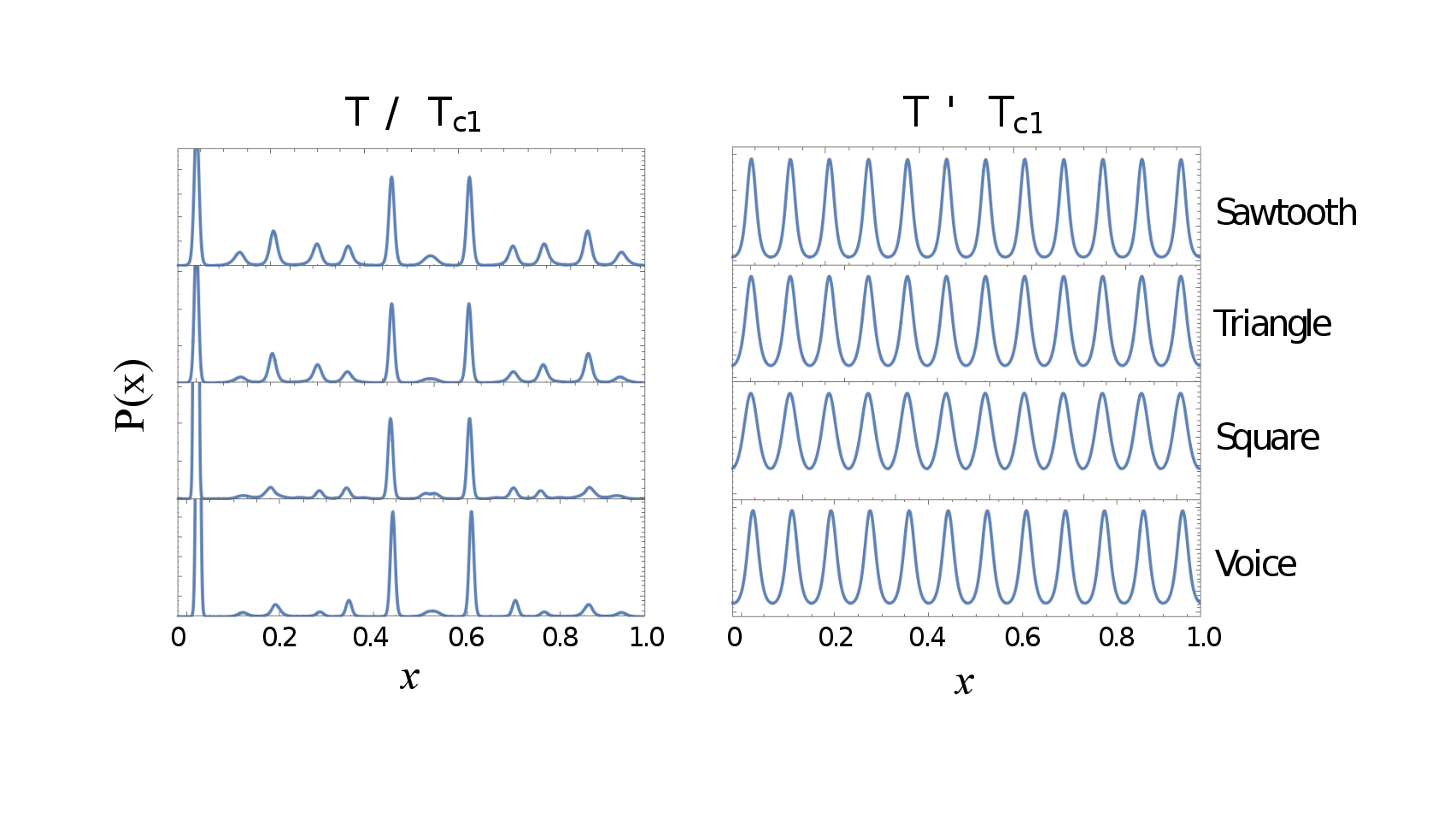}
\end{center}
\caption{Calculated $P(x)$ for four different harmonic timbres at two temperatures, one slightly below critical temperature $T_\textrm{c1}$ (left), and another slightly above $T_\textrm{c1}$ (right). From top to bottom, the timbres used were sawtooth, triangle, square, and human voice.}
\label{fig:Timbre_comparison}
\end{figure}

%Figure \ref{fig:Timbre_comparison} shows the largest negative Fourier coefficient of $D_P(x)$ as a function of $w_c$ for each of these timbres as well as the original sawtooth for comparison, and is thus a prediction of the periodicity of $P(x)$ between critical temperatures $T_1$ and $T_2$.  
In fig \ref{fig:Timbre_comparison} we show the calculated $P(x)$ for all four timbres at two temperatures, one slightly below critical temperature $T_\textrm{c1}$, and another slightly above $T_\textrm{c1}$. We can see from these comparisons that the model is fundamentally robust to a variety of integer harmonic timbres. However, we note that the transition rates around the critical temperatures can vary significantly with timbre, and if more than 10 harmonics are used in the calculation of $D_P(x)$, we do start to see some significant qualitative departures from the results shown in Fig. \ref{fig:Timbre_comparison}.

\subsection*{Relaxing the summation constraint}
We next investigate what happens with this model if we relax the constraint described in Eq(\ref{dp}). Note that since we are summing over positive and negative octaves, this definition of $D_P(x)$ is symmetric about the midpoint of its domain, $x=0.5$. That is, $D_P(x)=D_P(1-x)$. For this test we simply set $D_P(x) = D_\textrm{tot}(x)$, which means the symmetry of $D_P(x)$ about the domain midpoint is now no longer present. Since we would no longer expect that the resulting $P(x)$ would necessarily be periodic over the octave, we extend the domain of our calculation arbitrarily to three octaves, in order to investigate whether the octave periodicity deteriorates. 

%(since these three octaves contain all of the harmonics used in the simulation)

\begin{figure}[h!]
\begin{center}
\includegraphics[width=3.5in]{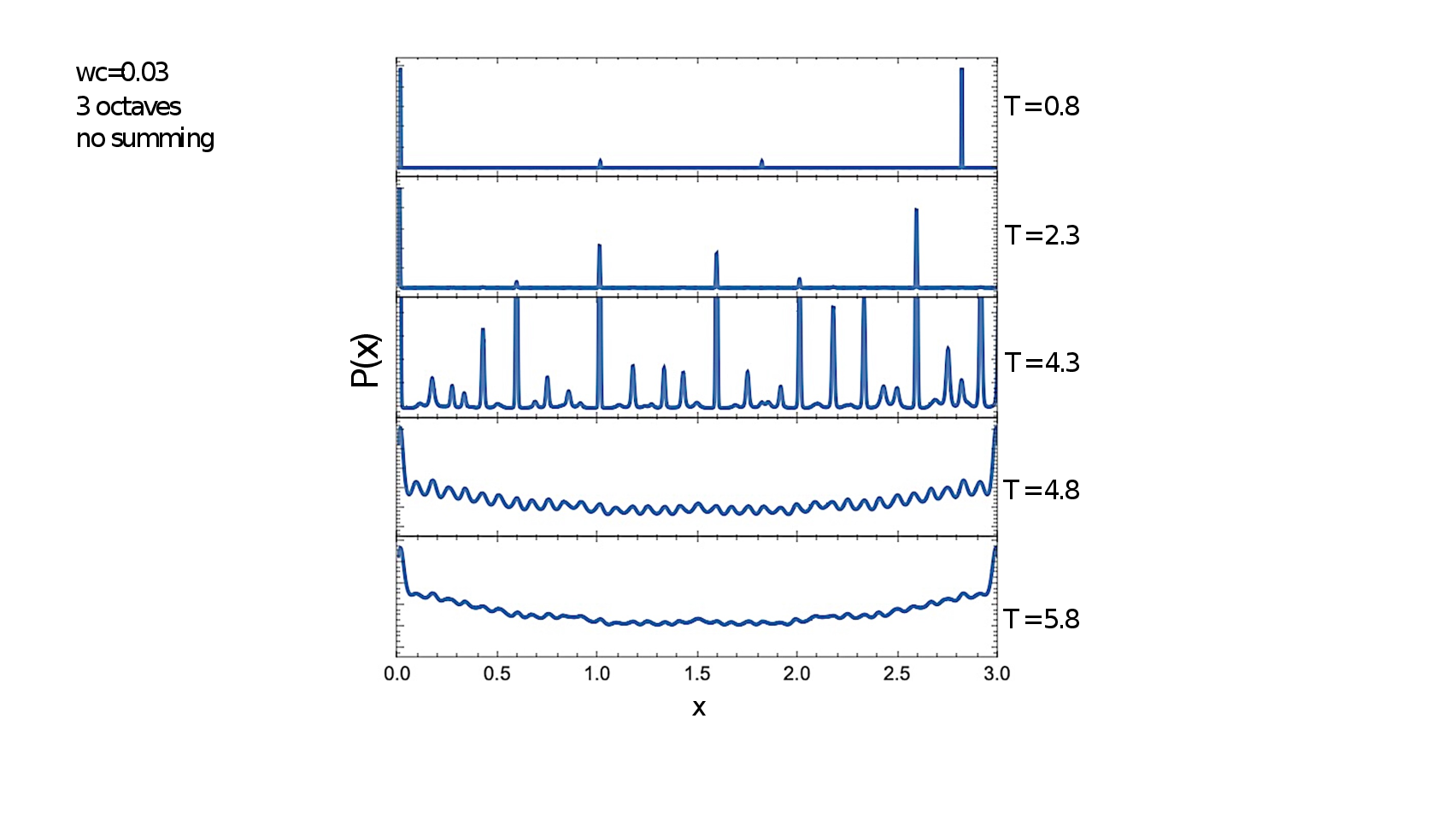}
\end{center}
\vspace*{-0.5cm}
\caption{P(x) over 3 octaves for the sawtooth waveform and $w_c = 0.03$. In this case, we do not impose periodicity or symmetry over the octaves, and instead simply define $D_P(x) = D_\textrm{tot}$.}
\label{fig:Sawtooth_no_summing}
\end{figure}

We once again return to the sawtooth waveform and perform a temperature scan over the three octave domain, setting $w_c = 0.03$. Fig. \ref{fig:Sawtooth_no_summing} shows a few of these temperatures. These results show that even if we relax the octave constraint in the definition of $D_P(x)$ and in the chosen domain size, we still see a tendency toward 12 pitches per octave. We can also see that $P(x)$ remains larger at the domain boundaries than in the middle,  even for high temperatures where we usually expect to find a flat $P(x)$. This is possibly because at those extreme $x$ values, pitches can no longer ``interact over the periodicity." That is, pitches just above $x=0$ simply do not interact with pitches just below $x=3$, and so less dissonance ``piles up" there.  By this logic, it also makes sense that $P(x)$ tends toward a global minimum near the middle of the domain. 

\begin{figure}[h!]
\begin{center}
\includegraphics[width=3.5in]{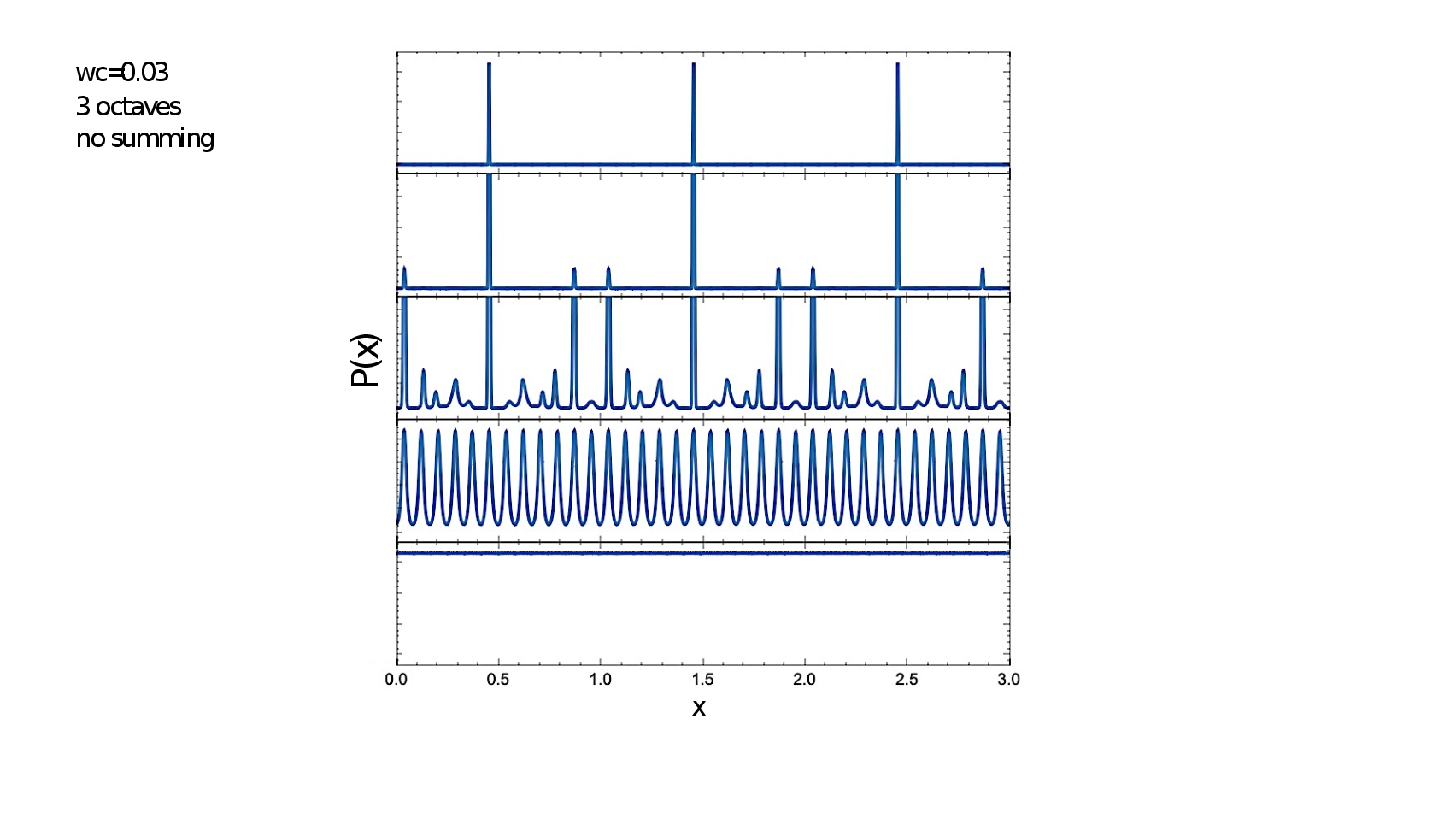}
\end{center}
\caption{Same conditions as Fig. \ref{fig:Sawtooth_no_summing}, but forcing $D_P(x)$ to be symmetric about the domain midpoint.}
\label{fig:Sawtooth_symmetric_summing}
\end{figure}

To justify this claim, we perform another test where we only force $D_P(x)$ to be symmetric about the midpoint of the domain by defining $$D_P(x) = D_\textrm{tot}(x) + D_\textrm{tot}(3-x)$$ over the domain from $x = 0$ to 3. The resulting $P(x)$ are shown in Fig. \ref{fig:Sawtooth_symmetric_summing}. Note that these are exactly the same results as the full octave summation (equation (4) in \cite{bib1}), shown in Fig. \ref{fig:6}, just repeated over three octaves. As predicted, by reintroducing only this particular $D_P(x)$ symmetry, we no longer see $P(x)$ increasing close to the domain boundaries. Both of these results seem to support that the octave simply emerges as ``natural" for an integer harmonic series, even when we do not sum over the octaves or impose a pure octave domain.

\subsection*{Gamelan}

We will now explore a special group of instruments from the Indonesian Gamelan tradition, which primarily features metallic percussion instruments, where unlike the case of instruments built to exploit standing waves in pipes and strings, non-integer harmonics arise. The bonang and saron are central to this ensemble, with the bonang made up of small gongs and the saron consisting of metal bars, both arranged horizontally on racks. Each gong or bar is precisely tuned to a specific pitch and played by striking it with either padded or hard mallets. The saron typically plays the ``balungan", or core melody, which serves as the foundation of a gamelan piece, while the bonang and other instruments add embellishments around this central structure. By contrast with traditional Western musical instruments, largely dominated by strings and pipes, the spectra of Gamelan instruments reveal peaks that do not occur in integer ratios, as the vibrating element is not a 1-D system, but a more complicated 2 and 3-D structure \cite{bib7}.

The prevailing musical scales of the Gamelan system are the five-pitch Slendro and the seven-pitch Pelog, both of which are significantly different in character to the Western chromatic divisions. Both of these scales are unevenly spaced within the span of a single octave, and the periodicity of the scale is usually not even marked by standard octave divisions (integer multiples, or 1200 cents), but rather an amount that is usually slightly larger (around $1210$ cents). See \cite{bib4}, page 213. In addition, the precise values of scale pitches sometimes vary from octave to octave. 

We wish to investigate whether our model can reproduce these scales. For the bonang harmonics, we have used
$$a_\textrm{bonang}(n) = (1.0, 1.52, 3.46, 3.92)$$  For the saron, 
$$a_\textrm{saron}(n) =(1.0, 2.34, 2.76, 4.75, 5.08, 5.91)$$
both of which are based on data reported in \cite{bib4}.
Since we do not wish to impose the octave as a natural musical division, we will once again allow our domain to span three octaves, and we will do no summing over the octave. In other words, once again we will simply set  $D_P(x) = D_\textrm{tot}(x)$.

\begin{figure}[h!]
\begin{center}
\includegraphics[width=5in]{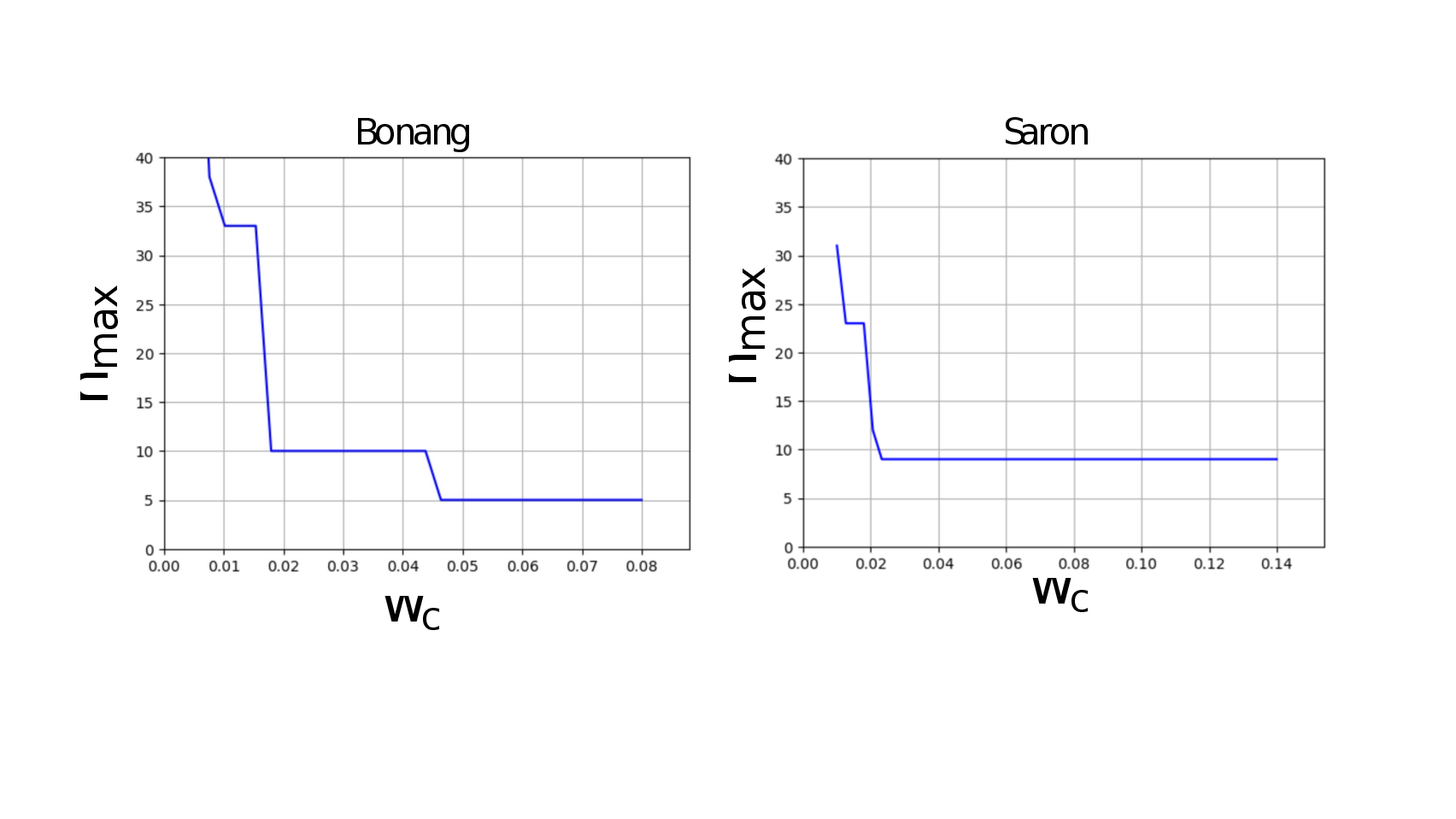}
\end{center}
\vspace*{-0.5cm}
\caption{Maximum negative Fourier coefficient, $n_\textrm{max}$, of $D_P(x)$, as a function of $w_c$, and thus a prediction of the periodicity of $P(x)$ over the octave for each harmonic series, bonang and saron.}
\label{fig:wc_scans}
\end{figure}

\begin{figure}[h!]
\begin{center}
\includegraphics[width=5in]{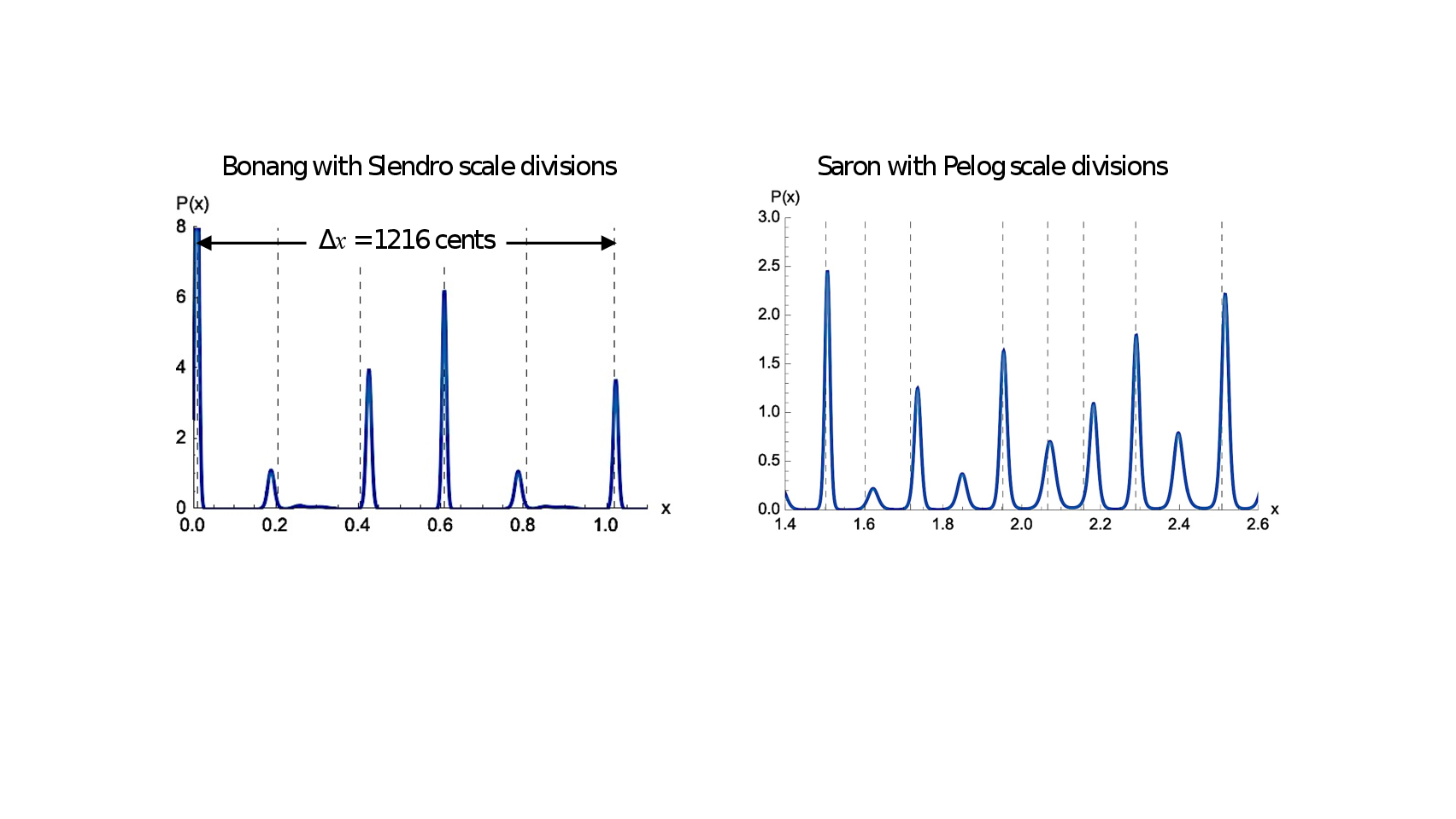}
\end{center}
\vspace*{-0.5cm}
\caption{Select regions of the three-octave P(x) for for each harmonic series, bonang and saron. Dashed lines indicate five pitch Slendro scale divisions on the left, and seven pitch Pelog scale divisions on the right.}
\label{fig:Slendro_Pelog}
\end{figure}

We can see in Fig. \ref{fig:wc_scans} the predicted maximum negative Fourier coefficient per octave, $n_\textrm{max}$ of $D_P(x)$ for both of these harmonic series as a function of $w_c$.  The Slendro 5-pitch scale is associated with the bonang, and so we will choose a $w_c$ value for which  $n_\textrm{max}  = 5$ ($w_c = 0.05$). The results of this $P(x)$ calculation for the bonang harmonic series are shown in Fig \ref{fig:Slendro_Pelog}. We can see clear evidence for something similar to the 5-pitch Slendro scale, the standard intervals of which we have indicated with dashed vertical lines. Interestingly, we can also see that the octave is stretched by about 18 cents, consistent with observations of typical Slendro tunings \cite{bib4}.

\vspace*{-0.5cm}

For the saron, which are typically tuned to the 7-pitch Pelog scale, we can see there are no values of $w_c$ that correspond to $n_\textrm{max}  = 7$.  Fig. \ref{fig:wc_scans}, right, shows that the majority of $w_c$ values predict $n_\textrm{max}  = 9$.  When we choose one of these values ($w_c = 0.04$), we do indeed see the emergence of 9 peaks in the $P(x)$ prediction, as shown in Fig. \ref{fig:Slendro_Pelog}. Typical tunings of the uneven Pelog 7-pitch scale are shown with dotted lines, and it seems that these lines align fairly well with a selected 7 of them. It tantalizingly suggests that the tuning of the Pelog scale instruments may be based on a subset of a chromatic 9-pitch scale, similar to the way that the major scale in Western harmony is a 7-pitch subset of a 12-pitch chromatic scale, as has been previously hypothesized by Braun \cite{bib22}.

%\begin{table}[!ht]
%\begin{adjustwidth}{-2.25in}{0in} % Comment out/remove adjustwidth environment if table fits in text column.
%\centering
%\caption{
%{\bf Table caption Nulla mi mi, venenatis sed ipsum varius, volutpat euismod diam.}}
%\begin{tabular}{|l+l|l|l|l|l|l|l|}
%\hline
%\multicolumn{4}{|l|}{\bf Heading1} & \multicolumn{4}{|l|}{\bf Heading2}\\ \thickhline
%$cell1 row1$ & cell2 row 1 & cell3 row 1 & cell4 row 1 & cell5 row 1 & cell6 row 1 & cell7 row 1 & cell8 row 1\\ \hline
%$cell1 row2$ & cell2 row 2 & cell3 row 2 & cell4 row 2 & cell5 row 2 & cell6 row 2 & cell7 row 2 & cell8 row 2\\ \hline
%$cell1 row3$ & cell2 row 3 & cell3 row 3 & cell4 row 3 & cell5 row 3 & cell6 row 3 & cell7 row 3 & cell8 row 3\\ \hline
%\end{tabular}
%\begin{flushleft} Table notes Phasellus venenatis, tortor nec vestibulum mattis, massa tortor interdum felis, nec pellentesque metus tortor nec nisl. Ut ornare mauris tellus, vel dapibus arcu suscipit sed.
%\end{flushleft}
%\label{table1}
%\end{adjustwidth}
%\end{table}

%\section*{Discussion}
%Nulla mi mi, venenatis sed ipsum varius, Table~\ref{table1} volutpat euismod diam. Proin rutrum vel massa non gravida. Quisque tempor sem et dignissim rutrum. Lorem ipsum dolor sit amet, consectetur adipiscing elit. Morbi at justo vitae nulla elementum commodo eu id massa. In vitae diam ac augue semper tincidunt eu ut eros. Fusce fringilla erat porttitor lectus cursus, vel sagittis arcu lobortis. Aliquam in enim semper, aliquam massa id, cursus neque. Praesent faucibus semper libero~\cite{bib3}.

\section*{Conclusion} In this paper, we have investigated in detail various aspects of the method proposed by  Berezovsky \cite{bib1} where the tools of statistical mechanics that are used to describe emergent order in phase transitions can also be used to show how harmony arises as an ordered phase of discrete pitches of sound. We have sought to clarify that the choice of timbre is important to determine the efficacy of the method (in \cite{bib1} only a sawtooth timbre is used without further comment or explanation for the choice). We have corrected some typos in \cite{bib1} that are relevant to the calculation of the Dissonance function, and given intuitive analytical arguments to  predict how the emergent ordered phase depends on largest Fourier coefficient of $D_{p}(x)$. We have further extended the model beyond the summation constraint over octaves, and have shown that the octave simply emerges as ``natural" for an integer harmonic series, even when we do not sum over the octaves or impose a pure octave domain, when considering instruments with integer harmonics such as pipes or strings.  We have also applied the method to see how the method can be applied to Gamelan musical systems that are explicitly non-periodic in octaves, and which therefore appear to fall beyond the purview of the method, and have shown it can still accurately capture these systems of tuning, allowing the possibility of interpreting that the 7-note pelog instruments could be understood as a subset of a larger 9 note scale, much in the same way that the major scale in Western intonation is a subset of a larger 12 note partition of the octave. This result is particularly interesting because traditionally, what students learn in a Harmony class is strictly speaking the harmonic style of 17th century European composers. It is our hope that this first paper allows us to bring attention to the fact the thermodynamic framework proposed in \cite{bib1} is much more robust and powerful than it may seem at first; it allows us to understand multiple tuning systems used across human history and culture as a natural outcome that only depends on the details with which said cultures perceive dissonance, and therefore brings all forms of music culture to have an equal seat at the table. After all, the results show they are in essence no different from the natural processes that give rise to order in chemistry and biology that have long been understood in terms of the minimization of a free energy.

% For more information, see \nameref{S1_Appendix}.

%\section*{Supporting information}

% Include only the SI item label in the paragraph heading. Use the \nameref{label} command to cite SI items in the text.
%\paragraph*{S1 Fig.}
%\label{S1_Fig}
%{\bf Bold the title sentence.} Add descriptive text after the title of the item (optional).

%\paragraph*{S1 Appendix.}
%\label{S1_Appendix}

%\paragraph*{S1 Table.}
%\label{S1_Table}
%{\bf Lorem ipsum.} Maecenas convallis mauris sit amet sem ultrices gravida. Etiam eget sapien nibh. Sed ac ipsum eget enim egestas ullamcorper nec euismod ligula. Curabitur fringilla pulvinar lectus consectetur pellentesque.

\section*{Acknowledgments} Luis Nasser gratefully acknowledges the support
from the NSF award PHY - 2110425. Xavier Hernandez acknowledges financial assistance CONAHCYT and DGAPA grant IN-102624.

\nolinenumbers

% Either type in your references using
% \begin{thebibliography}{}
% \bibitem{}
% Text
% \end{thebibliography}
%
% or
%
% Compile your BiBTeX database using our plos2015.bst
% style file and paste the contents of your .bbl file
% here. See http://journals.plos.org/plosone/s/latex for
% step-by-step instructions.
%

\end{document}